# Chapter 3

# Thermal conductivity of glasses and disordered crystals

A.I. Krivchikov[a] and A. Jeżowski[b]

[a]B.Verkin Institute for Low Temperature Physics and Engineering of the NAS of Ukraine, 47 Nauky Ave., Kharkiv, 61103, Ukraine

krivchikov@ilt.kharkov.ua

[b]Institute of Low Temperature and Structure Research, PAS, 2 Okolna Str., 50-422 Wrocław, Poland

A.Jezowski@int.pan.wroc.pl

Among the many physical properties, the amorphous state manifests itself in the most spectacular way in heat transport. Anomalously low thermal conductivity, its low-temperature dependence as a function of temperature, the presence of a plateau of thermal conductivity, all are definitely different when compared with the ordered state of crystals. For this reason, the microscopic mechanisms from which these surprising behaviors may emerge have been the subject of intensive theoretical research and even more extensive experimental research. These investigations have led to the detection of a huge range of disordered materials with somewhat similar properties, which have become a rich base for various physical interpretations. In this chapter we present the last 50 years of the history of experimental research together with the key theoretical physics scenarios.

## 1. Introduction

This chapter deals with experimental studies of the thermal conductivity $\kappa$ of solids, which is characterized by a unique temperature dependence of the same type, the so-called glass-like thermal conductivity, in a wide range from ultralow to high (1000 K) temperatures. The turning point in the understanding of glassy behavior of the thermal conductivity was the pioneering work[1] of Zeller and Pohl in 1971. For the first time they undertook a joint measurement of low-temperature thermal properties (heat capacity and thermal conductivity) of several structural glasses such as $SiO_2$, $GeO_2$, multi-component structural glass Pyrex, and aluminogermanate, as well as amorphous Se, in a temperature range from 0.05 to 100 K. A comparative analysis of the obtained results and literature data for a large number of amorphous substances of various nature allowed the authors to reach the fundamental conclusion that the low-temperature behavior of the thermal properties of amorphous substances is rather insensitive to the chemical composition of the material and is quite different from the behavior of the corresponding crystals, and that cannot therefore be explained in the framework of collective excitations of a phonon gas in a crystal lattice. They found a direct correlation between the additional linear-in-temperature contribution to the heat capacity and the thermal conductivity quadratic dependence $\kappa \sim T^2$. This allowed them to formulate the problem of low-temperature anomalies of disordered solids, which was the inspiration for plenty of new experimental research not only of thermal properties, but also of others, among them dielectric, acoustic, structural and mechanical ones. Schematic dependences of reduced heat capacity $C/T^3$ and glass-like thermal conductivity of a disordered solid on its reduced temperature, around the temperature of the boson peak (see Chapters 2, 8 and 9) $T_{max}$, are shown in Fig. 1.

## 2. Low-temperature thermal conductivity data

To date, thermal conductivity has been investigated in more than 100 amorphous solids and in more than 80 crystalline substances, in which glass-like behavior takes place. Studies performed after the work of Zeller and Pohl can be divided into several stages, that we separate by decades in the following.

## 2.1. *Studies in the decade 1971-1980*

The initial period (from 1971 to 1980) was primarily associated with experimental work on the measurement of the thermal conductivity of typical amorphous substances and glasses: **polymers**[1-12]; **chalcogenide glasses**[13,14]; **structural glasses**[14-19]; **metallic glasses**[20-22]; **amorphous solids**[23-25]; **radiation damaged crystals**[26]; **mixtures of glass oxides**[27].

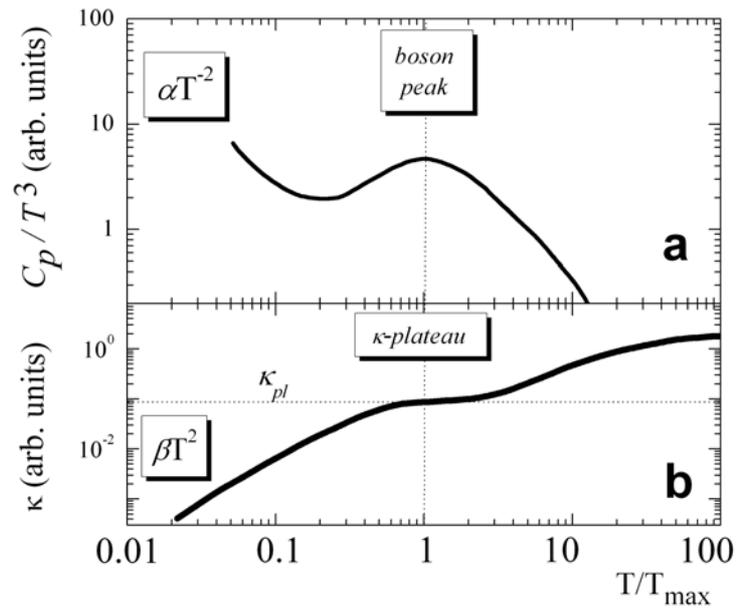

Fig. 1. Schematic dependences of reduced heat capacity $C/T^3$ (a) and glass-like thermal conductivity (b) of a disordered solid on its reduced temperature $T/T_{max}$, below and above the temperature of the boson peak $T_{max}$. The thermal conductivity quadratic factor is $\beta$ and the plateau thermal conductivity is $\kappa_{pl}$.

The above cited works show that thermal conductivity of amorphous solids (obtained in various ways) exhibits some universal behavior, which is due to the interaction (resonant scattering) of acoustic excitations with a set of localized low-energy excitations according to the well-known phenomenological tunneling model of Anderson, Halperin and Varma[28], and of Phillips[29], at least at the lowest temperatures (see Section 3.1. below).

In the same period, in the first review of the thermal conductivity of polymers[12], it was noted that *"the temperature dependence for all amorphous polymers is approximately equal in magnitude and characterized by a $T^2$ dependence below 0.5 K, a plateau region between 5 and 15 K and a slow increase at yet higher temperatures"*.

## 2.2. *Studies in the decade 1981-1990*

In the 1980's decade – the second important period – an exciting event emerged, namely, the detection of glass-like behavior not only in new amorphous substances, but also in crystalline structures, such as ferroelectrics, ionic conductors, crystals with orientational disorder, as well as in radiation-damaged crystals. Many experimental investigations revealed new features of glass-like behavior in the $\kappa$-plateau and high-temperature increase in molecular glasses, mixture of glass oxides, chalcogenide glasses, disordered crystals, as well as in a variety of metal glasses and in many modern –at that time– polymers and structural glasses. The search also began for explanation of which external factors (pressure, electric field, radiation, variation in composition of polymeric and quasicrystalline materials), affect the glass-like behavior in a wide temperature range. Information had appeared on the nature of the thermal activation driving of the high-temperature dependence of thermal conductivity in clathrate hydrates[30]. Among others, the following works were published in this decade: ***ferroelectric crystals***[31-37]; $Pb(Sc_{0.5}Nb_{0.5})O_3$ under electric field[38]; tourmaline in pyroelectric state[39]; $PbZrO_3$ in pyroelectric state, antiferroelectric[40]; ***ionic conductors and ceramics***[31,41-44]; ***orientational glasses***[45-49]; ***mixtures of glass oxides***[50-54]; ***molecular glasses***[55,56]; ***clathrate compounds***[30,57-59]; ***quasicrystals***[60]; ***radiation damaged solids***[61,62]; ***amorphous solids***[63-67]; ***structural glasses***[53,56,68,69];

SiO$_2$ porous[70,71]; SiO$_2$ aerogel[72]; SiO$_2$+Ne[73,74]; ***chalcogenide glasses***[75]; ***polymers***[56,76-85]; electron-irradiated epoxy resin[86], epoxy resin under pressure[87]; ***metallic glasses***[88-97].

In the same period, a large number of reviews[75,85,98-109] were published, in which the results of many experimental investigations of different properties, including thermal conductivity of glasses, disordered crystals and glassy crystals were presented. In Ref. 102, new data on the glass-like behavior of thermal conductivity in metallic and insulating glasses, and in crystalline disordered solids, were discussed. The glass-like thermal conductivity of complex inclusion crystals with orientational disorder, clathrate hydrates at high temperatures and under pressure, was discussed in Ref. 103. Another review[104] stressed that the low temperature glassy properties of disordered crystals exhibit the same anomalous behavior as found in amorphous solids. However, these did not provide a definitive answer to the main problem. *"There is no final answer to the main problem. The major problem is of course the microscopic nature of the tunneling entity."*[105] In Ref. 106 there were reviewed some inorganic ferroelectrics, antiferroelectrics and pyroelectrics, which display glass-like thermal properties at low and ultralow temperatures. It was found[107] that glass-like behavior of thermal conductivity in ferroelectric single crystals of relaxor-type is sensitive to an electric field.

A review by Phillips[108] covered a wide range of experimental and theoretical studies of two-level or tunneling states in glasses and concluded: *"A convincing microscopic description of a tunneling state in any amorphous material remains a problem to tax the solid state physicist."* Cahill and Pohl[75] showed that in amorphous and highly-disordered solids, the picture of quantized vibrations exhibits serious deficiencies at high frequencies. Greig[85] noted that there is not any '*universal*' thermal conductivity behavior in semi crystalline polymers. Also, several general reviews of experimental works were published, concerning anomalous very-low-energy properties of nonmetallic solids and their theoretical interpretation.[98-102,105,110]

### 2.3. *Studies in the decade 1991-2000*

In the following stage (during the 1990's decade), it was first discovered that a complex polycrystalline filled skutterudite[111] $Ir_4NdGe_3Sb_9$ containing weakly bound atoms in some crystal nodes, and almost-crystalline three-component metal structures called quasicrystals, show glasslike behavior. Pioneering works were done in the study of the extreme effects of pressure, radiation, and electric field on thermal conductivity of glasses, polymers and crystals. At that time, evidence emerged that the thermal conductivity of porous $SiO_2$ aerogel anomalously depends on its density. It was also revealed that thermal conductivity of metallic glasses at high temperatures presents the same regularity as structural glasses and amorphous films. Crystalline clathrate compounds, significantly different in chemical bonding, display the usual glassy behavior. Further, it was found that orientationally-disordered crystals of both inorganic and organic substances, as well as molecular glasses, present all the characteristic features of the glass-like behavior of thermal conductivity. Studies of a large number of both single crystals and mixed crystals revealed only new quantitative features of the glass-like behavior of the universal temperature dependence of thermal conductivity in a wide range of temperatures. For example thermal conductivity of disordered solids slightly increases with increasing density. Data on the small Bridgman parameter $g \approx 2 \div 3$ for a group of polymer substances were obtained[112-115] and later confirmed[116,117].

The following works were published in this decade: ***quasicrystals***[118-120]; ***skutterudites***[111]; ***amorphous solids***[121,122], a-Si porous[123]; ***radiation damaged crystals:*** neutron-irradiated $C$[124]; ***structural glasses:*** $SiO_2$ film[125,126], $Al_2O_3$ film[127], aerogel $SiO_2$[128,129], densified $SiO_2$[130], $SiO_2$ under pressure[131]; ***mixtures of glass oxides***[132,133]; ***polymers***[112-115,134-146]; ***molecular glasses***[147]; ***orientational glasses***[134,149,150]; ***clathrate compounds***[151-157]; ***metallic glasses***[158,159]; ***ferroelectric crystals***[160-165].

As in the previous period, several general reviews of experimental works were published[132,166-178]. These reviews were related to anomalous very-low-energy properties of metallic and nonmetallic glass and disordered crystals, either under normal condition or under pressure. They also dealt with theoretical interpretation within the framework of the soft-potential model[171,179,180], which is able to explain the glassy anomalies in the specific heat and the thermal conductivity in a much wider low-temperature range. Other works considered the phonon-glass electron-crystal concept[175], or a lower limit to the thermal conductivity of solids[181]. This experimentally observed lower limit to the thermal conductivity of solids was ascribed to the minimum thermal conductivity earlier proposed by Einstein, in which the elastic energy propagates in a random walk among the atoms which are vibrating with random phases.

## 2.4. *Studies in the decade 2001-2010*

In the fourth stage, i.e. in the first decade of the new millennium, considerable efforts were directed to experimental and theoretical studies of crystals with a different kind of dynamic, static, orientational, positional, polaronic, and other kinds of disorder (see Fig. 2). At this stage, researchers revealed various new crystals with both weak and strong interatomic forces, in which glass-like behavior of thermal conductivity takes place. It became clear that the $\kappa$-plateau can manifest itself not only in the low-temperature region (4—10 K), but also in the medium-temperature region (20—50 K) in the case of metallic glasses, quasicrystals, in some semiconductor clathrate crystals and in some ferroelectrics and ceramics. In the temperature range above the "plateau", the thermal conductivity of quasicrystals is described by the Arrhenius type equation. At that time, it has been suggested that avoiding the crossing between heat carrying acoustic modes and the low lying optical modes[182-184] can influence glass-like behavior of thermal properties. The following works were published in this decade: ***clathrate compounds***[185-2023,245]; ***quasicrystals***[203-208]; ***orientational glasses***[209-211]; ***skutterudites and other rattling structures***[212-214]; ***ferroelectric crystals***[215,216,244]; ***ceramics***[217]; ***molecular glasses***[213,217-220]; ***amorphous solids***[221-225]; ***structural glasses***[226-229]; ***chalcogenide glasses***[230,231]; ***polymers***[119,232-235], ***films***[236]; ***metallic glasses***[237-239]; ***others***: ZrHfTiNiSn[240], zeolite NaX[241], $(La_{1-x}Yb_x)_2Zr_2O_7$[242].

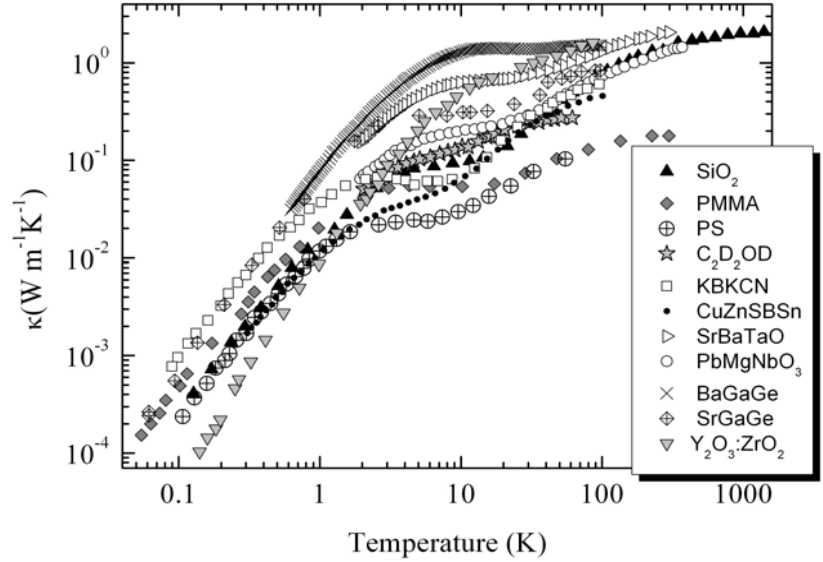

Fig.2. Glass-like thermal conductivity of different glasses and disordered crystals: $SiO_2$ (Refs. 16,56); polymers PMMA, PS (Ref. 15); molecular glass $C_2D_5OD$ (Ref. 243); orientational glass $(KBr)_{0.75}(CN)_{0.25}$ (Ref. 47); ferroelectric crystal PMN (Ref. 244), SrBaTaO (Ref. 296); clathrate compounds SGG (Ref. 153), BGG (Ref. 245) and CuZnSbSn (Ref. 246); single crystal $Y_2O_3:ZrO_2$ (Ref. 247).

Also several general reviews of experimental and theoretical works were published[182,214,248-254]. In Ref. 250, attention was drawn to "… *the fact that the low-energy excitations in numerous disordered crystals resemble very closely those found in amorphous solids also strongly suggests a picture in which lack of long-range order per se is not the cause of the low-energy excitations. But in that case, what is important?*" Using the example of over 60 different compositions, it was shown that $T^2$ dependence of thermal conductivity is universal and is dominated by phonons that are resonantly scattered by the low-energy excitations in the frame of the tunneling model.

## 2.5. *Studies in the decade 2011-2020*

In the fifth stage, which comprises this last decade, the study of glass-like behavior in new organic and inorganic crystals of various thermoelectric materials, deuterated substances, porous amorphous films and glasses, unusual ceramics, and ferroelectrics is actively continuing. In this period, new representatives appeared in the discussed large family of "glass-like" behavior, for example, BSA, myoglobin proteins, $Tb_2Ti_2O_7$ spin-liquid, disordered layered structures and organic metal anisotropic single crystals, cyanate ether resin, HMS ribbon, nanofiber and nanocrystals. It has been established that deuteration leads to an increase in the high-temperature thermal conductivity of clathrate hydrates[154], amorphous water[255] and molecular glasses[243]. A relevant line of current research was noted in the review of Chang and Zhao[256] where the relationship between anharmonicity and low thermal conductivity in thermoelectrics was summarized. Several strategies which yield anharmonicity were also suggested, including lone pair electron, resonant bonding and the rattling model found in clathrates and skutterudites with cage-like structure.

The following main works were published in this decade: ***biomaterials***[257]; ***clathrate compounds***[246,258-264]; ***quasicrystals***[265]; ***orientational glasses***[243,266-268]; ***molecular glasses***[243,269-272]; ***amorphous solids***[253,273-277]; ***structural glasses***[278-280]; ***polymers***[281-292]; ***metallic glasses***[293,294]; ***ferroelectric crystals***[295-298]; ***ceramics***[247,299-304]; ***others***: layered crystal structure of $NaZn_4Sb_3$[305], $Tb_2Ti_2O_7$ spin liquid[306], silica zeolite[307], Si nanowires[308], boron carbide nanowires[309], lead halide perovskite nanowires[310], amorphous nanowires[311,312], SiN films[313], a-$Si_3N_4$ nanowires[314], Si nanocrystals[315], HMS ribbon[316], θ-(BEDT-TTF)$_2$MZn(SCN)$_4$[317], $Ln_3NbO_7$(Ln=Dy,Y,Er,Yb)[318], $CH_3NH_3PbBr_3$ nanowire[319], hydroxyapatite nanoparticles[320], ***nanostructures***[321-325].

Also several general reviews of experimental and theoretical works have been published[288,315,326-335]. In Refs. 326,328 important correlations between the properties at very low temperatures and at moderately low temperatures in glasses are discussed in the frame of soft-potential model and other theoretical approaches. In Ref. 327 the authors discussed recent advances in materials engineering to control thermal conductivity and stated that "… *good progress has been made with materials possessing rattling atoms or complex unit cells*". A similar statement can be made on the basis of another review[329], which presents a large variety of properties of phonon-glass electron-crystal thermoelectric clathrates. The thermal properties under pressure and the transition behavior of several host-guest inclusion compounds (urea, thiourea, Dianin's compound, clathrate hydrates and hydroquinone) have been reviewed in Ref. 330. There, the unusual glass-like thermal conductivity of inclusion compounds was found to be of great technological and fundamental importance to seek for improved thermoelectrical materials, despite the fact that the origin of their glass-like thermal conductivity is not understood. In another review[331], it can be read: "*Recently, however, a growing number of studies have re-examined the thermal properties of amorphous semiconductors, such as amorphous Si. These studies, which included both computational and experimental work, have revealed that phonon transport in amorphous materials is perhaps more complicated than previously thought*". On the other hand, De Angelis et al.[333] have reviewed the theory and computational modeling approaches that can be applied to disordered materials and discussed corresponding experimental techniques. They concluded that "*… important questions lay on the horizon for better understanding thermal transport when disorder is present and tremendous progress is possible in the next few decades.*"

## 3. Theoretical descriptions

Disordered solids haven't translation symmetry and therefore Peierls's theoretical picture that heat carriers are the crystalline excitation of propagating vibrational waves seems to break. To explain this anomalous behavior of an amorphous solid, various phenomenological models were developed that introduced a number of low-energy excitations: tunneling two-level systems (TLS), relaxation systems and low-frequency quasilocal vibrational states, in addition to acoustic continuous-medium excitations. The temperature dependence of the thermal conductivity is linked to the nature of the energy transport which is quantified by ballistic propagation and thermal diffusivity of collective excitations in a disordered solid. According to the Standard Tunneling Model and the Soft-Potential Model, in ballistic regime or in a quasiparticle picture, continuous-medium acoustic excitations are strongly resonant scattered by TLS excitations and quasi-local vibrations. A harmonic theory of thermal transport[336] in glasses was proposed by Allen and Feldman in year 1989: an additional thermal diffusive regime or wave interference picture was implemented where heat is carried by new carriers, the coupling of high-energy vibration modes being named diffusons and locons. Allen-Feldman approaches, however, do not always provide physical insight into how the modes are actually contributing to the thermal conductivity.[336,337] The thermal conductivity of complex ordered crystals displays a glass-like low-temperature behavior[338-340] (the universal $T^2$ dependence predicted by the tunneling model) and a glass-like high-temperature behavior[219,260,266,316,341-343] (decay milder than the universal $T^{-1}$ dependence predicted by the phonon gas model). It should be noted one review[333] that outlined theoretical computational modern approaches, summarized the progress in understanding materials with disorder, and highlighted open questions.

### 3.1. *Standard Tunneling Model*

The Standard Tunneling Model was proposed by Anderson, Halperin and Varma[28], and by Phillips[29] in 1972 to describe many aspects of the universal low temperature properties of amorphous solids. Two level tunneling states (TLS) with a constant density of states as defects of an elastic continuum were postulated. For an elastic continuum such as a glass, the thermal conductivity $\kappa(T)$ is given by the standard expression obtained from the well-known phonon-gas kinetic equation, and using the Debye approximation for the density of states of the sound waves transporting heat. The thermal conductivity quadratic dependence $\kappa = \beta T^2$ is evaluated on the assumption of strongly resonant scattering of continuous-medium excitations by TLS excitations. One obtains: [28,29,250]

$$\kappa(T) = \frac{k_B^3}{6\pi\hbar^2}\left(\sum_\alpha \frac{\rho s_\alpha}{\overline{P}\gamma_\alpha^2}\right) T^2 = \frac{k_B^3}{6\pi\hbar^2}\left(\sum_\alpha \frac{1}{C_\alpha s_\alpha}\right) T^2, \qquad (1)$$

with

$$C_\alpha = \frac{\overline{P}\gamma_\alpha^2}{\rho s_\alpha^2} \qquad (2)$$

where $\alpha$ is the polarization of phonons, $\overline{P}$ is the spectral density of the tunneling states, and $\gamma$ is the sound excitation-TLS coupling. $\rho$ and $s$, respectively, represent the mass density and the properly sound velocity.

### 3.2. *Soft-Potential Model*

For temperatures between 1 and 10 K, the anomalous behavior of thermal conductivity that manifests itself by the appearance of a temperature-independent plateau, as well as a concomitant increase in heat capacity well above that corresponding to the Debye model, was the subject of conflicting explanations. The explanations have been now rationalized on phenomenological grounds by means of a generalization of the tunneling model known as the soft-potential model (SPM) [180,344,345] and also as soft-mode model [346].

In the model of soft potentials, glass is considered as an elastic medium with specific structural defects, the particle motion of which can be not only harmonic, but also anharmonic. It has been shown to be able to account for the plateau of the thermal conductivity on the basis of the assumption of resonant scattering of sound waves from localized low-frequency vibrations. The soft-potential model describing elementary excitations in a wide range of energies introduces a characteristic energy $W$ (often expressed in K), which assigns the energy scale in the classification of elementary excitations in the harmonic soft potential. According to Ref. 180, $\kappa(T)$ is described by a universal function in the representation of a normalized temperature (reduced temperature) $u = T/W$, which can be read

$$\kappa(T) = \frac{6k_B^3}{\pi\hbar^2}\frac{W^2}{\overline{C}\cdot s}\frac{T^2}{1.1W^2 + 0.7TW + 3T^2} \quad , \qquad (3)$$

The dimensionless parameter $\bar{C}$ takes on a fairly narrow interval $10^{-3} \div 10^{-4}$ and characterizes the strength of the interaction of TLS with sound excitations at averaged sound velocity *s*.

The expression (3) allows us to relate the observed three characteristics −the plateau in thermal conductivity ($\kappa_{pl}$), the thermal conductivity quadratic factor ($\beta$), a characteristic energy ($W$)− to each other in the following form:

$$\frac{\kappa_{pl}}{\beta W^2} \approx \frac{1}{3} \quad . \quad (4)$$

The ratio (4) is in good agreement with experimental data for disordered solids of different nature in the range of parameters $\kappa_{pl} \approx 10^{-3} \div 1$ Wm$^{-1}$K$^{-1}$, $\beta \approx 0.01 \div 0.1$ Wm$^{-1}$K$^{-3}$ and $W \approx 0.5 \div 15$ K.

## 4. Recapitulation

The large amount of materials described above, in which glass-like thermal conductivity is observed, raises the question of finding common physical patterns, and also raises the question of the factors influencing the character of this property.

A lot of work has clearly demonstrated a correlation[31] between the maximum temperature $T_{max}$ (boson peak) in the reduced heat capacity $C/T^3$ and temperature "glass-like" plateau in the thermal conductivity.[15,31,179,326,347-350,361]. This unambiguous correlation was observed both in joint and independent measurements of the heat capacity and thermal conductivity of glasses and amorphous solids of different nature, as well as in complex crystals with disorder, for example: ***amorphous solids***[1,24,222,224]; ***chalcogenide glasses***[54,75,231,351,352];

***clathrate compounds***[151,245,246,188-189,199,201-202,258,259,262,264]; ***ferroelectric crystals***[31,36,38,164,244,217]; ***ionic conductors***[31,247,300]; ***metallic glasses***[237,238]; ***mixture of glass oxides***[353]; ***molecular glasses***[217,269,270]; ***orientational glasses***[45,47]; ***radiation solids***[19,354]; ***polymers***[78,79,140,143,355]; (also under pressure)[87]; ***structural glasses***[1,356]; ***other materials***[240].

The correlation Boson peak–plateau persists after external and internal effects on a solid when varying factors such as pressure[87,130,140,141,144,145], annealing[73], electric field[161,162], neutrons irradiation[26,62] or electrons irradiation[19]. The position and size of the plateau depends on the porosity of the solid[129,227,356] and on the concentration of the solvent in the two-component solid solution[50,53,97,199,216,353,357].

However, it should be noted that there is also a skeptical view on the existence of a direct correlation between the specific features of heat capacity and thermal conductivity, which can be expressed in the following quote: "*Our most important conclusion is that the "reincrease" of thermal conductivity above the plateau region is attributable to heat carried by "diffuson" modes in much the way imagined by Birch, Clark, and Kittel, and that the plateau is a simple crossover region, not requiring any new physics to explain. In particular, we believe that "excess modes" (also known as a "Boson peak") is not a necessary ingredient to explain the plateau. Amorphous silicon seems to lack these "excess modes" but still to have a plateau.*"[358] But the following fact seem to contradict the above statements. Namely, the correlation Boson peak–plateau can be clearly seen if we represent the dependence of thermal conductivity $\kappa$ on the temperature normalized to the temperature of the maximum of the boson peak $T_{max}$ (see Fig. 3). In the $T \approx T_{max}$ region, the $\kappa(T/T_{max})$ exhibits systematically a plateau. This circumstance can be used for a new definition of the term "glass-like thermal conductivity".

The glass-like thermal conductivity can be characterized by the following four features:

(i) $\kappa(T)$ is proportional to $\beta T^{2-\delta}$ ($\delta = 0 \div 0.2$) at $T \ll T_{max}$. (ii) There is a plateau region the plateau thermal conductivity $\kappa_{pl}(T) \approx const$ near $T_{max}$.

(iii) $\kappa(T)$ has a rising inflection point above the plateau from $T > T_{max}$.
(iv) A saturating regime where $\kappa(T)$ is independent of $T$ for $T > 10 T_{max}$.

It is to be emphasized that the $\kappa$-plateau is very sensitive to external factors. From a large set of experimental results, we can distinguish in principle the following effects that affect the thermal conductivity:

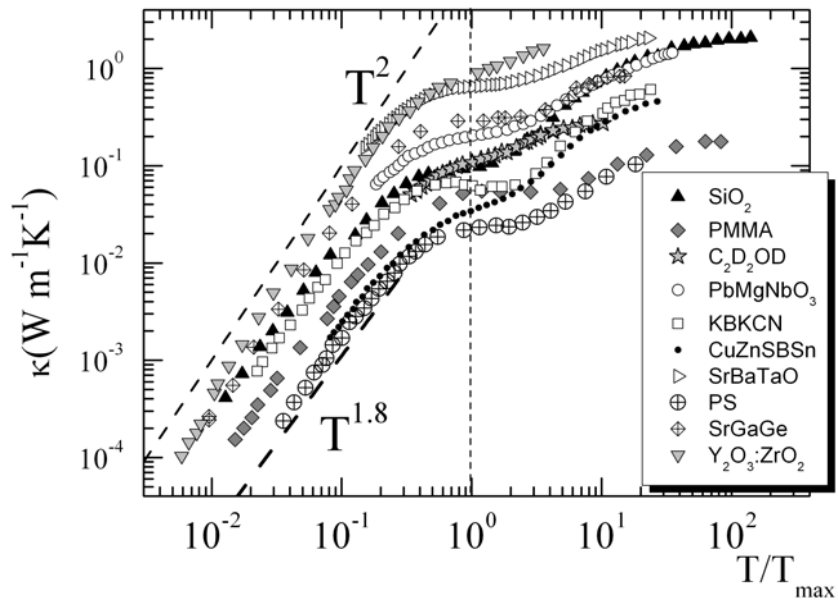

Fig. 3. Glass-like temperature thermal conductivity depending on scaled temperature by the $T_{max}$ for glasses and crystalline materials (see the symbols in the caption of Fig. 2).

- Low temperature $T^{2-\delta}$ effect – most of the data in the literature in the temperature range 0.1–1 K report a $\delta > 0$, but for $T < 0.1$ K $\delta \approx 0$: $SiO_2$[17,228], $SBN$[173], $B_2O_3$[228], $PVC$[232], $Zr_{52.5}Ti_5Cu_{17.9}Ni_{14.6}Al_{10}$[293], Polyimide[292]. A nearly linear specific heat and a thermal conductivity roughly proportional to $T^2$ ($\kappa(T) = \beta T^2$) are common features of glass-like behavior. As thermal conductivity is concerned, not only is this universality qualitative but also quantitative[250]. For many different glassy solids the universal $T^2$ temperature dependence of the thermal conductivity is in reasonable agreement with the prediction of the tunneling model. The values of the coefficient $\beta$ for the $T^2$ dependence are located in a rather confined interval the thermal conductivity quadratic factor $\beta \approx 0.01 \div 0.1$ Wm$^{-1}$K$^{-3}$. The experimental results of thermal conductance of amorphous nanowires between 0.05 K and 5K show a uniform quadratic dependence on the temperature.[311,312]

- The effect of pressure – increase of heat conductivity in the $\kappa$-plateau area with increase in external pressure, as observed for epoxy resin[87], SBN[160], epoxy resin[140], PC[145], PS[141].

- The effect of irradiation – a change in the thermal conductivity at the $\kappa$-plateau with increasing dose: falling for epoxy resin[88]; rising for SiO$_2$[19,27].

- Electric field effect (poling) – increase of thermal conductivity in the $\kappa$-plateau region with an increase in the magnitude of the electric field: Pb(Sc$_{0.5}$Nb$_{0.5}$)O$_3$, SBN[35,38,162], Pb(Mg$_{1/3}$Nb$_{2/3}$)O$_3$[162,359].

- The "shoulder effect" – a transformation of the plateau into a curve with a slightly increasing slope is observed in two-component substances: 0.9SiO$_2$·0.1GeO$_2$ and 0.925SiO$_2$·0.075Ti$_2$O[133], (1-$x$)B$_2$O$_3$·$x$Na$_2$O[52], Ge$_x$As$_{0.4-x}$S$_{0.6}$[231], Fe$_{0.4}$Ni$_{0.4}$P$_{0.14}$B$_{0.06}$[91], and PMN$_{1-x}$[PbTiO$_3$]$_x$[215].

- Porosity effect – decrease of heat conductivity in the $\kappa$-plateau area with an increase in porosity: see Refs. [11,70,129,227,278,356].

- Cross-link density effect – increase of thermal conductivity in the $\kappa$-plateau area with a decrease of cross-link density: epoxy resin[78,79,81,136,357], PS[84].

- Annealing effect – a change of thermal conductivity in the $\kappa$-plateau under the action of annealing: falling for PC[360]; rising for SiO$_2$[73].

- The effect of deuteration – an increase in the value of high-temperature thermal conductivity in deuterated substances (above the plateau): THF + 17D$_2$O[154], C$_2$D$_5$OD[243], a-ice[255].

This huge amount of work on glass-like behavior over the last 50 years shows how fascinating it was and still is the problem. Many common features and analogies have been found, e.g. the resonant phonon scattering by localized excitations and the concept of a lower limit to the thermal conductivity. But, despite the fact that so many interesting results, including thermal conductivity data, have been collected, the question about the nature of observed anomalies in so significantly different objects remains open.